\documentstyle[12pt,a4nodot]{article}     % Paper without sections
\newcommand{\be}{\begin{equation}}
\newcommand{\ee}{\end{equation}}
\newcommand{\bea}{\begin{eqnarray}}
\newcommand{\eea}{\end{eqnarray}}
\newcommand{\bean}{\begin{eqnarray*}}
\newcommand{\eean}{\end{eqnarray*}}

\begin{document}
%\begin{titlepage}
%\begin{flushright} RAL-93-051\\
%\end{flushright}
%\vskip 1cm
%\begin{center}
{\bf\large HADRONS AND GLUE AT A  TAU-CHARM FACTORY}
\vskip 1cm
\noindent  { \it  F E Close}\\
Rutherford Appleton Laboratory,\\
Chilton, Didcot, Oxon,
OX11 OQX, England.\\
%\end{center}
\begin{abstract}
I discuss the special opportunities that a Tau-Charm Factory offers for
identify
 ing
gluonic excitations, hybrid charmonium and other exotic hadronic states.
\end{abstract}
%\end{titlepage}

\section{Introduction}

The QCD Lagrangian contains quarks and gluons and the successes of
perturbative QCD confirm their existence.  The behaviour of QCD in the strongly

interacting regime, ``non perturbative QCD", is less well understood
theoretically.
There is copious data in the form of hadron spectroscopy and decays which needs
``only" to be interpreted  in order to establish the  properties  of the full
nonperturbative
theory.  It is clear that $J=1/2$ quarks and antiquarks are excitable degrees
of

freedom in the strongly interacting  regime  and there are also manifestations
o
 f
QCD
in the spectrum (e.g. hyperfine splittings between $Jp=0-$ and $1-$ mesons)
but a
great unsolved question is whether gluonic degrees of freedom can be excited
in the non-perturbative limit and be manifested as a new spectroscopy.  If this
spectroscopy exists then its systematics may reveal insights into the nature of
confinement and other aspects of strong interaction dynamics.  There are two
parts to the general question:  is glue excited\\
i) on its own, forming ``glueballs", $G$;\\ ii) in the presence of quarks,
formi
 ng
``hybrids", $H$.

Theoretical prejudice and models tends to suggest that the primitive $G$ and
$H_q$ (hybrids involving $u,d,s$ flavours) exist in the 1-2 GeV mass region
[1,2,3]  and
so may be mixed  with, and confused with, the $q\bar{q}$ spectrum.  Heavy
hybrids, $H_Q$, are expected to occur on a mass scale [4,5,6]
$$
m(H_Q) \simeq m(Q\bar{Q}) +0(1\; GeV).
$$
Thus  for the charmed hybrids one anticipates $H_c$ at or around the $DD*$
threshold.  Given that the conventional charmonium spectroscopy is
rather well understood below 4 GeV, and that the region at and immediately
above threshold
is a high priority at a  Tau-Charm Factory  (TCF), there is a good prospect
that

gluonic charmonium may
be isolated by a dedicated survey of the region from threshold up to 5 GeV of
c.m.   energy.  Indeed, if at a TCF one is unable to determine whether hybrid
or

gluonium
exist, then we may have to accept that this is a very difficult experimental
question.

\section{Light Hadrons}
\subsection{At all energies}

A strategy for isolating glueballs or other exotics involves first identifying
a
 ll of
the expected $q\bar{q}$ states.  Those with $C=+$ can be produced  in
$\gamma\gamma$ collisions.  $\gamma\gamma$ is useful because the
$\gamma$ couples to the flavour and spin of the constituents in
well-understood ways.   There are detailed predictions
 for the relative couplings of states in a given supermultiplet and also
selection rules [7] that can help to complete the $q\bar{q}$ spectroscopy.
Stud
 y is
needed to determine the  extent to which $\gamma\gamma$ physics can
parasite on
dedicated studies at various energies.

\subsection{At the $\psi$}

$\psi \rightarrow \gamma X$ has been one of the most productive processes in
modern strong interaction studies.  It still offers us more.  Questions
include\
 \
i)   $\psi \rightarrow \gamma (C=+$) complements the
$\gamma\gamma\rightarrow  (C=+)$.  This can help identify gluonic versus
$q\bar{q}$ states by their ``stickiness" (a preference for the former process).
 [8]\\
ii)   $\psi \rightarrow \gamma \eta$ (1440): is this a single state and if so
wh
 at is
it (see $\psi \rightarrow \gamma \gamma V$ below).\\
iii) $\psi\rightarrow\gamma \theta$ (1700): is this state $0+$ or $2+$ - or
both?  If $2+$ is present,
what are the relative strengths of its production in different helicity states?
[9]  Can this distinguish gluonic from $q\bar{q}$ or $K*K*$? [10]\\
iv) $\psi\rightarrow\gamma \eta\pi\pi$.  When looking for the enigmatic
scalars $f_0$ (975), $a_0$ (980) [11, 12] or other states in the $\eta\pi\pi$
sy
 stem,
this
``clean" channel has advantages over $p\bar{p}\rightarrow\pi\pi$ as in the
Crystal Barrel.\\
v) $\psi\rightarrow\gamma  X, X \rightarrow\rho$ or $ \gamma V$ where
$V=\rho\omega$ or $
\phi$ makes use of the ideal flavours of the vector mesons to enable a flavour
tag
to be made of the state $X$.  An $X (s\bar{s})$ decays preferentially to $
\gamma\phi$  rather
than to $\gamma\rho$ or $ \gamma\omega$; $X(u\bar{u}+d\bar{d}$)
$\rightarrow
\gamma\rho, \gamma\omega$ rather than $\gamma \phi$.  A glueball or
flavour singlet produces all in the ratio
$$
\Gamma (G\rightarrow \gamma\rho : \gamma\omega : \gamma\phi) =
9:1:2
$$
Such studies are at the limit b of present statistics for $X=\eta $ (1440) and
b
 eyond
them for $\theta$ (1700); of the order of 1000 events per state could be
extract
 ed
from
$109$ total $\psi$  decays at TCF.  In addition to isolating the flavour
conten
 ts of
the
$C=+$ mesons, for $J_X\geq 2$ the relative helicity amplitudes for
$X\rightarrow\gamma V$ may also be measured and give information on the
internal structure of $X$.

\subsection{At the $\psi$ (3684)}

In a single week of running at $\psi$ (3684) one may obtain $2\times 106 \chi$
in $\psi\prime\rightarrow \gamma\chi$.  This is already a factor of 10 higher
then the present world total.  Thus, running at $\psi$ (3684) makes the TCF
a {\bf $\chi$-factory}.

i) $\chi_J\rightarrow$ hadrons:  This will provide much new information as
data on $\chi$ decays are very sparse.  When looking for resonances in
sequential decays, e.g. $\chi_J\rightarrow A+\pi \rightarrow (n \pi)+\pi $, the
knowledge of $J$ helps to constrain the analysis.  This may make significant
inroads into understanding the spectroscopy of light hadrons.

ii) $J$-filter: Particular values of $J$ may be advantageous for studying
specif
 ic
$J{PC}$ of high interest.  Examples include the $S$-waves,
$$
\chi_0\rightarrow (f_0(975) \; or\; a_0(980)) + 0{++}\; ;
$$
$$
\chi_0\rightarrow \pi + 1{-+}\; ,
$$
and so $\chi_0$ provides a gateway to the $0{++}$ system and $\chi_1$ directly
accesses the $I=1$ exotic partial wave $J{PC}=1{-+}$.  Any resonance in this
wave
simply cannot be  $q\bar{q}$.  Light hybrids, $H_q$, may occur in the 1.5-2 GeV
region [2,3,6] and appear  in
$$
\chi_1 \rightarrow \pi H_1\rightarrow \pi (\pi f_1).
$$

iii) Flavour filter: These complement the $\psi\rightarrow VX, TX$ decays but
with different quantum numbers accessible in the final state.  In particular
the

flavours of the $f_0$ (975) may be probed in the relative strengths of [12]
$$
\chi_\rightarrow f_0 f_2 (1270): a_0 a_2 (1320) : f_0 f (1525)
$$
which emphasise respectively the $f_0 (n\bar{n})$ : $a_0  (n\bar{n})$ and $f_0
(n\bar{n})$.

\section{Charmonium $c\bar{c}$}

The theory is ``clean" for the narrow states below $D\bar{D}$ threshold.  The
strategy must be to complete the
spectroscopy of narrow states and to clarify the situation above $DD$ and
$DD{**}$
thresholds.

We have heard about the possibility to isolate narrow states in
$p\bar{p}\rightarrow\chi$.  The advantage in $p\bar{p}$ is in the resolution by
which widths can be measured; a disadvantage is that it can be like hunting for
 a
needle in a haystack.  The missing (or recently discovered $1P_1$) states [13]
 can
be
accessed at TCF as follows
\bea
\eta\prime_c (21S_0) &:& \quad \psi\prime \rightarrow \gamma
\eta\prime_c\nonumber \\
1{+-}  (1P_1)& :& \quad \psi* \rightarrow \eta 1P_1\; at\; \sqrt{s} > \; 4
\
 ;
GeV\nonumber \\
2{--}  (3D_2) &: &\quad \psi* \rightarrow \eta (2{--})\; at\; \sqrt{s} > \;
 4.5
\;
GeV\nonumber \\
2{-+}  (1D_2) &:& \quad \psi* \rightarrow \gamma (2{-+})\; at\; \sqrt{s} =
\;
4.03 \; GeV\; peak\nonumber
\eea
In the last example, if the B.R. = $0(10{-3}$) then we anticipate $103$
events
  in a
single day; which illustrates the promise    of TCF.

It is important to run on the structures already identified at 4.04, 4.16 and
4
 .42
GeV to
identify the branching fractions to hadronic final states (essentially {\bf
noth
 ing}
is known here).  It will
be interesting to measure the relative abundance of $DD: DD*: D*D* (D=0-,
D*=1-$) as well as the $D{**} (0+, 1+, 2+$) in the higher mass bumps.

According to heavy quark effective theory,  in the limit in which  fine and
hyperfine mass splittings between the $D_J$ states vanish, one expects  various
characteristic production ratios to occur in the
continuum [14,15,16].  These may be measured to test HQET,
probe non-trivial configurations within the $D_J$ states, help to isolate
missin
 g
$D_J$ states (e.g. $0+, 1+_{1/2}$) and to identify the internal structure of
$\psi*$ states above 4 GeV.

Thus, as one example, one expects either in the continuum or on a
$\psi*(3S_1$) state to find the ratios
$$
D\bar{D}: \bar{D}D* +\bar{D}*D: D*\bar{D}* = 1:4:7
$$
(up to $0(\alpha_s$) corrections) [14],
whereas for $\psi*(3D_1$) one expects  [15]
$$
\psi(3D_1)\rightarrow DD:DD*+D*D : D*D* = 1:1:4
$$
Interesting deviations from these ratios may be anticipated near to threshold
for
$$
\frac{\Delta m(D*-D)}{\sqrt{s-4M2}}  \simeq 0(1)
$$
At the $\psi$ (4.04) for example there is the possibility that, if this state
is

$33S_1$ say, the nodes in the $33S_1$ wavefunction may lead to a significant
distortion of the above ratios.  As $D,D*,F,F*$ are produced with different
values of momentum it may be possible to map out momentum space nodes for
the $\psi*$.

Ref (17) has argued that the apparent large ratio in favour of $D*D*$ at the
$\psi$ (4.04) may be an example of this.  Analogous arguments previously
applied to the
$\psi$ (4.4) [18], when combined with the now known  values of the $\Delta m
(D{**}-D$),
suggest that $D{**}_2$ may be copiously produced at the $\psi$ (4.42).  It is
important to study the structures in the $e+e-$ total cross-section above the
charm threshold to see what the relative $D_J$ content within each is,  both to
clarify which values of energy are optimal for producing particular states
$D_J$

and to distinguish   true $\psi* c\bar{c}$ resonances from   threshold
effects in  $e+e-\rightarrow D_{J_1}D_{J_2}$.

\section{Beyond  $Q\bar{Q} : Q\bar{Q}g$ Hybrids}

A summary of theoretical predictions for hybrid charmonium is:\\
i) {\bf Mass} 4.4 $\pm$ 0.4 GeV (or 4.2 $\pm$ 0.2 GeV if less conservative).
Th
 is
range spans lattice QCD [5], MIT Bag model [2,4] and flux tube models [3,6].
\\ ii) {\bf Exotic quantum numbers}:  All models include $J{PC} = 1{-+}$ in
th
 e
lowest mass supermultiplet of hybrids.  Flux tube models also expect  the
exotic
 s
$0{+-}$ and $
2{+-}$ to be low lying.\\ iii) {\bf Decay modes} : If $M>$ 4.3 GeV the
dominant

decays will be to $DD{**}$ (i.e. $S+P$ states) in which case widths should  be
typically hadronic (perhaps 100 MeV if phase space allows). [3,6,19]

If $M<$ 4.3 GeV widths will be reduced $(perhaps \leq$ 10 MeV) and the
significant decays may include
cascades into charmonium, $H_c\rightarrow\chi_c$ + hadrons, with some
$DD$ and $DD*$.  The relative importance of these modes  is
model dependent.\\
iv)  {\bf Production}.  All models anticipate $J{PC}=1{--} H_c$ in  the
lowest supermultiplet and so they will be accessible directly in $e+e-
\rightarrow H_c$.  In flux-tube models the flux has a transverse excitation
around the $Q\bar{Q}$ axis and the $Q\bar{Q}$ are  themselves in  an orbitally
excited state with a $P$-wave centrifugal barrier.  Thus one may
anticipate that
$$
\Gamma{ee} (\psi 3S_1) >   \Gamma{ee} (\psi_g) >
\Gamma{ee} (\psi 3D_1)
$$
In the MIT bag approach, $0(\alpha_s)$ mixing between $3S_1$ and $c\bar{c} g$
will drive the leptonic width and a similar pattern may be anticipated.    For
l
 ight
quarks, where such suppression is not dramatic, one may anticipate a
reasonable
$\Gamma{ee} (V_g) $ but this is confounded by  the difficulty of
unambiguously
identifying the states among the detailed $\bar{q}$ spectroscopy.  By contrast,
  for
$b\bar{b}$ systems  the spectroscopy is clean but  $\Gamma{ee}
(\Upsilon_g$)  is more strongly  suppressed (e.g. there is no evidence for any
$\Upsilon (3D_1$)
states in $e+e-$ annihilation).  The Catch-22 is that clean spectroscopy
equat
 es
with minimal mixing (good news) but minimal mixing equates with a small
$\Gamma{ee} $ for all but $3S_1$ states.  This argument suggests that
$c\bar{c}$  may provide on balance  a useful compromise: non-trivial mixing
(e.g.
$3D_1$ states do  appear in $e+e-$ annihilation) but with a
manageably clear spectroscopy.

For a vector hybrid  below  4.4 GeV one may anticipate as order of
magnitudes estimates  $\Gamma_T = 0$ (10 MeV), $\Gamma{ee}$ 0(0.1 keV).
This
would give a local peak in $R\equiv\sigma (e+e-\rightarrow$ hadrons)/
$\sigma (e+e-\rightarrow \mu+\mu-$) of $\Delta R$ 0(1-2) and $>$ 10
events per second at TCF.  Allowing an order of magnitude for conservatism, a
scan of $R$ from threshold to the highest machine energy should be sensitive to
$$
\Delta R = 0.1, \quad \Delta E = 5\; MeV.
$$
This requires about  2000 events per energy setting.   For $\ell = 10{33}
cm{-2} sec{-1}$,
yielding 10 events/sec, one could scan this entire energy range in a week.  A
monochromator,
enabling a very fine scan with high statistics, is desirable to reveal, and to
clarify the nature of  detailed structures in this region of resonances and
open  thresholds.

If above 4.4 GeV, $\psi_g, \psi(3S_1)$ and $\psi(3D_1)$ all have couplings to
$DD{**}$ ($S$ and $P$ state charmed mesons).  This will cause mixing and also
coupling to $e+e-$.  Thus one anticipates an excess of $J{PC}=1{--}\psi$
sta
 tes
relative to that in the naive quark model.  A strategy will include first
determining the spectrum and then identifying the pattern of $0{-+}$ or
$1{-+}
 $
states nearby.

Estimates of hyperfine mass shifts for hybrids suggest that $0{-+}$ and
$1{-+}
 $
(exotic) are, respectively, of order 50 MeV and 25 MeV lower in mass than the
$1{--}$ hybrid.  For $\psi(3S_1)$ there should be an accompanying $0{-+}$
state but no $1{-+}$.  For $\psi(3D_1)$ there is neither a $0{-+}$ nor  a
$1
 {-+}$
partner.  These three distinct patterns are illustrated in fig 1.

The matrix element for the radiative transition
$\psi_g\rightarrow\gamma\eta_{cg}$ is essentially the same as that for
$\psi\rightarrow \gamma\eta_c$.  The relative rates will depend on the
mass splitting
$$
\Gamma(\psi_g\rightarrow\gamma \eta_{cg})\sim
\Gamma(\psi\rightarrow\gamma \eta_c) \left(\frac{\Delta M (\psi_g-
\eta_{cg})}{\Delta M(\psi-\eta_c)} \right)3
$$
If  $M(\eta_{cg} <$ 4.3 GeV) and its total width is thereby only a few MeV, it
m
 ay
be
possible to detect the 20-50 MeV photons which accompany the transitions to
$0{-+}$
and $1{-+}$.  However if $M(\eta_{cg} >$ 4.3 GeV), the strong decays to
$DD{**}$ will reduce the radiative branching ratio.

Instead of directly forming a  $\psi_H$  hybrid it may be possible to produce
them  [20]
from $e+e-$ at $\sqrt{s}\sim$ 5 GeV.  In the continuum process   $e+e-
\rightarrow c\bar{c}$,    the separating quarks will be accompanied by a flux
tube that in general will contain a superposition of modes including
excitations

or ``hybrid" states.  The dynamics of the flux tube may cause the cascade
preferentially to contain $S$-wave mesons in a relative $P$-wave e.g.
$\psi_H\rightarrow\eta\psi*$ (where $\psi*$ could be $\psi_H$ or normal
$\psi$).
The $\psi*$ state in turn may cascade into $\psi$ via
$\psi*\rightarrow\eta\psi$ or $\psi*\rightarrow\pi\pi\psi$.  Thus running
$e+e-$ at high energy and triggering on $\psi$ in the final state may reveal
$\psi*$ in the ($\eta\psi)$ or $ (\pi\pi \psi$) invariant mass.  These
$\psi*$

will
include states that have already shown up in the direct  process $e+e-
\rightarrow\psi*$;
it will be particularly interesting if $\psi*$ states  which show up
prominentl
 y
in this ``contimuum
cascade"   are suppressed in $e+e-\rightarrow\psi*$.  Thus in
summary: seek hybrid states in
$$
e+e-\rightarrow\eta\psi_H \rightarrow \eta (\eta\psi).
$$
The qualitative pattern of such transitions is illustrated in fig 2.

Triggering on $\psi$ may enable access to other exotic states that are not
direc
 tly
produced in $e+e-$ annihilation.  For example, there is  considerable
discussion about the possible existence of ``meson-molecules"; the   $a_0$
(980)

and
$f_0$ (975) being $K\bar{K}$ states and a possible $KK*$ state occurring in
the

$1{++}$ partial wave around 1400 MeV.  These suggest the question: is there
molecular charmonium?  Indeed, the apparent enhancement of
$D*\bar{D}*$ around 4.04 GeV has been suggested as such. [21]

A clear example of such states would be if $I=1$ states occur, thus
$$
\pmatrix{M_0\cr M_1\cr} = \frac{1}{\sqrt{2}} (D0D0 \pm D+D-)
$$
To the extent that $e+e-\rightarrow c\bar{c}$ initially, then only $I=0$
direc
 t
channel bound states are formed:
$$
e+e-\rightarrow \not{\rightarrow} (D\bar{D})_1
$$

However, an $I=1\; D\bar{D}$ molecule  state may be accessed by cascading, and
in turn, would be revealed by its
subsequent cascade into $\psi\pi$:
$$
e+e-\rightarrow (c\bar{c})   \rightarrow\pi  (D\bar{D})_1
\rightarrow\pi  (\pi\psi)
$$
This can be performed at any incident $e+e-$ energy.  An enhancement in the
$\pi\psi$ state would signal either isospin violating decays of charmonium or
the existence of isovector charmonium.  Either result involves  important
information.

\section{Which energies are optimal for hadron physics?}

As an initial proposal for discussion I list some ``interesting" energies, the
``factory" physics immediately relevant, their associated relevance to the
questions raised in this talk and guides as to the typical events in nominal
running time.

The $\psi$ and $\psi\prime$ can provide increases in world statistics even if
TCF operates at $10{32}$ luminosity.  Clearly early running of the machine
will

concentrate on these two peaks.  A fine detail scanning of $R$, with ability to
discriminate $\Delta R$ = 0.1 on $\Delta E$ = 5 MeV can be achieved in a week
at $10{33}$, in 3 months at $10{32}$.  There should be dedicated runs at 4.03
 (see
also CP violation review here), 4.16 ($D_s$ factory), 4.42 ($\Lambda_c$
factory)

and at 4.8 - 5 GeV ($\Xi_c$ and $\Omega_c$).  The first three bumps should be
clarified as resonant, multiple resonances or as threshold effects; their
branch
 ing
ratios must be established.  In addition higher spin $D{**}_J$ should be
sought

just above their production thresholds where the cross section peaks.  At the
to
 p
energy one should see if $e+e-\rightarrow D f_0 \bar{D}$ or $D a_0\bar{D}$ is
significant (as may occur in Gribov's theory of confinement [12]).

The $\psi$ (3772) is a natural $D$ factory.  There is a tantalising glitch in
th
 ree
separate experiments which has  $\Delta R\simeq $ 0.2 at 3.9 GeV.  This seems
too sharp to be a threshold; it may be a clue to vector hybrid and is at least
a
  place
to start.

I am indebted to my collaborator Ted Barnes for discussions and comments in
preparing this contribution.


\begin{thebibliography}{35}
\bibitem{ } T. Burnett and S. Sharpe, Ann. Rev. Nucl. Part. Sci. {\bf 40}, 327
(1990).
\bibitem{ }  T. Barnes and F.E. Close, Phys. Lett. {\bf 116B}, 365 (1982);\\
M. Chanowitz and S. Sharpe, Nucl. Phys. {\bf B222}, 211 (1983);\\
T. Barnes, F.E. Close and F. de Viron, Nucl. Phys. {\bf B224}, 241 (1983).
\bibitem{ } N. Isgur and J. Paton, Phys. Rev. {\bf D31}, 2910 (1985).
\bibitem{ } P. Hasenfratz, R. Horgan, J. Kuti and J. Richard, Phys. Lett. {\bf
9
 5B},
299 (1980).
\bibitem{ } C. Michael and S. Perantonis, Nucl. Phys. {\bf B347}, 854 (1990).
\bibitem{ } T. Barnes, F.E. Close and E. Swanson, RAL-93- (in preparation)
\bibitem{ } T. Barnes, F.E. Close and Z-P. Li, Phys. Rev. {\bf D43}, 2161
(1991)
 ;
  E. Ackleh, T. Barnes and  F.E. Close, Phys. Rev. {\bf D46}, 2257 (1992).
\bibitem{ } M. Chanowitz, Nucl. Phys. {\bf A527}, 61 (1991).
\bibitem{ } F.E. Close and Z-P. Li,  Phys. Rev. Lett. {\bf 66}, 3109 (1991).
\bibitem{ } F.E. Close and Z-P. Li,  Z. Physik  {\bf C54}, 147 (1992);\\
K Dooley, E. Swanson and T. Barnes, Phys. Lett. {\bf B275}, 478 (1992).
\bibitem{ } R.L. Jaffe,   Phys. Rev.  {\bf D15}, 267 (1977); {\bf D17}, 1444
(19
 78);\\
N. Isgur and J. Weinstein,  Phys. Rev. Lett.  {\bf 48}, 569 (1982);  Phys. Rev.
  {\bf
D27}, 588 (1983);  Phys. Rev.  {\bf D41}, 2236 (1990);\\
 F.E. Close, N. Isgur  and S. Kumano,  Nucl. Phys.  {\bf B389}, 513  (19913.
\bibitem{ } F.E. Close, Y. Dokshitzer, V.N. Gribov, V. Khoze and M. Ryskin,
RAL-93-049.
\bibitem{ } K. Seth, these proceedings
\bibitem{ } A. de Rujula, H. Georgi and S. Glashow, Phys. Rev. Lett. {\bf 37},
3
 98
(1976);\\ T. Mannel, W. Roberts and Z. Ryzak, Nucl. Phys. {\bf B368}, 204
(1992)
 .
\bibitem{ } F.E. Close and G.J. Gounaris,  Phys.  Lett. (in press).
\bibitem{ } P. Bialas, F.E. Close, J. Korner and K. Zalewski,  RAL-93- (in
prep)
 .
\bibitem{ } A Le Yaouanc, L Oliver, O. Pene and J. Raynal, Phys. Lett. {\bf
71B}
 , 57
(1977).
\bibitem{ } A Le Yaouanc, L Oliver, O. Pene and J. Raynal, Phys. Lett. {\bf
71B}
 ,
397 (1977).
\bibitem{ } F.E. Close and H.J. Lipkin,  Phys.  Lett. {\bf B196}, 245 (1987);\\
 H.J.
Lipkin,  Phys.  Lett. {\bf B219}, 99 (1989);\\ F. Iddir et al, Phys.  Lett.
{\bf
  B205},
564 (1988).
\bibitem{ } We thank D. Bugg for this suggestion.
\bibitem{ } See also N. Tornqvist, Phys. Rev. Lett. {\bf 67}, 556 (1991); and
P.
  Ball,
H. Dosch and M. Shifman, Phys. Rev. {\bf D47}, 4077 (1993).
\end{thebibliography}
\end{document}